\documentclass[prb,twocolumn,floatfix,superscriptaddress, longbibliography]{revtex4-1}
\usepackage[pdftex,plainpages=false,colorlinks=true,linkcolor=blue, citecolor=blue, urlcolor=blue]{hyperref}
\usepackage{amsfonts}
\usepackage{amsmath}
\usepackage{amssymb}
\usepackage{graphicx}
\usepackage{natbib}
\usepackage{color}
\usepackage{placeins}
\usepackage{physics}

\begin{document}
 
\title{Prediction of anomalies in the velocity of sound for the pseudogap of hole-doped cuprates}
\author{C. Walsh}
\affiliation{Department of Physics, Royal Holloway, University of London, Egham, Surrey, UK, TW20 0EX}
\author{M. Charlebois}
\affiliation{D\'epartement de Chimie, Biochimie et Physique, Institut de Recherche sur l’Hydrog\`ene, Universit\'e du Qu\'ebec \`a Trois-Rivi\`eres, Trois-Rivi\`eres, Qu\'ebec, Canada G9A 5H7}
\author{P. S\'emon}
\affiliation{D\'epartement de physique, Institut quantique \& RQMP, Universit\'e de Sherbrooke, Sherbrooke, Qu\'ebec, Canada J1K 2R1}
\author{G. Sordi}
\email[corresponding author: ]{giovanni.sordi@rhul.ac.uk}
\affiliation{Department of Physics, Royal Holloway, University of London, Egham, Surrey, UK, TW20 0EX}
\author{A.-M. S. Tremblay}
\affiliation{D\'epartement de physique, Institut quantique \& RQMP, Universit\'e de Sherbrooke, Sherbrooke, Qu\'ebec, Canada J1K 2R1}
\date{\today}

\begin{abstract}
We predict sound anomalies at the doping $\delta_{p}$ where the pseudogap ends in the normal state of hole-doped cuprates. Our prediction is based on the two-dimensional compressible Hubbard model using cluster dynamical mean-field theory. We find sharp anomalies (dips) in the velocity of sound as a function of doping and interaction. These dips are a signature of supercritical phenomena, stemming from an electronic transition without symmetry breaking below the superconducting dome. If experimentally verified, these signatures may help to solve the fundamental question of the nature of the pseudogap -- pinpointing its origin as due to Mott physics and resulting short-range correlations. 
\end{abstract}
 
\maketitle

\section{Introduction} 

Upon decreasing the temperature near half-filling, cuprate superconductors reach a temperature $T^*$ where spin susceptibility and photoemission experiments show features that suggest a loss in the density of states. This has been associated to a so-called pseudogap phase, with $T^*$ a crossover rather than a sharp phase transition~\cite{Alloul2013, annurev2019}. 
In addition, below $T^*$ as defined above, broken symmetry states are observed by some probes~\cite{annurev2019, keimerRev}. At even lower temperature, superconductivity and charge-density wave phases appear~\cite{annurev2019, keimerRev}. 
On the other hand, in 2018, experiments revealed that thermodynamic properties show signatures of a phase transition upon crossing the critical doping $\delta_{p}$ where the pseudogap ends at low temperature, such as a sharp peak in the specific heat~\cite{Michon:Cv2018}. 
Even more puzzling, this thermodynamic signature of a transition occurs without a detectable diverging correlation length associated to a broken symmetry state~\cite{Michon:Cv2018, annurev2019}.

The two-dimensional (2D) Hubbard model in the doped Mott insulator regime~\cite{Alloul2013} captures these aforementioned key features that are consistent with the observed pseudogap -- namely (i) that the pseudogap is a crossover across $T^*$ and (ii) there is a phase transition across $\delta_{p}$ at low $T$, (iii) without the need of a broken symmetry state, (iv) which can however appear within the pseudogap phase. 
In particular, studies on this model based on cluster extensions of dynamical mean-field theory~\cite{rmp} revealed that the phase transition across $\delta_{p}$ is a metal to metal first-order transition without symmetry breaking~\cite{sht, sht2}, due to Mott physics and short range correlations. 
This transition ends at finite doping and finite temperature in a second order critical point, from which supercritical crossovers emerge~\cite{ssht, sshtRHO}. This theoretical framework resolves the paradox of the thermodynamic anomalies at the endpoint of the pseudogap (e.g. the peak  in the specific heat across $\delta_{p}$) without the need for broken symmetry states, as discussed in Refs.~\cite{Giovanni:PRBcv,Alexis:2019}. 

In this article we provide a novel and crucial {\it prediction} of this theoretical framework for ultrasound experiments in hole-doped cuprates. The propagation of sound is a powerful tool to characterise phase transitions in solids because it is a sensitive way of probing the thermodynamics of a system. We shall prove that the velocity of sound has signatures of a phase transition across $\delta_{p}$ and of a crossover across $T^*$. Crucially, if confirmed by experiments, this will provide further decisive evidence that the pseudogap arises from Mott physics and short-range correlations. 

Recent advances in ultrasound techniques should enable the testing of our prediction. Indeed, the last few years have witnessed a renewed attention in using ultrasounds to probe electronic phase transitions in strongly correlated electron systems, and experimental findings motivate new theoretical investigations.
For example, an experiment~\cite{Shekhter2013} in the cuprate YBa$_2$Cu$_3$O$_{6+x}$ (YBCO) found a break in the slope of the velocity of sound at the onset of the pseudogap, suggestive of a phase transition, although this interpretation has been challenged in Ref.~\cite{Cooper2014}. In another cuprate, La$_{2-x}$Sr$_x$CuO$_4$ (LSCO), ultrasound anomalies were connected to the coupling between the lattice and the spin glass~\cite{Frachet:NatPHys2020, Frachet:PRB2021}. 
In other correlated electron systems, ultrasounds were used to place constraints on the order parameter symmetry of the hidden order in URu$_2$Si$_2$~\cite{Ghosh:SciAdv2020} and in the superconducting state of Sr$_2$RuO$_4$~\cite{Benhabib:NatPhys2020, Ghosh:NatPhys2020}. Anomalies in the velocity of sound have been used to describe the crossover emerging from the Mott transition in V$_2$O$_3$~\cite{Populoh2011} and organic superconductors~\cite{Fournier2003, Poirier:PRB2011}. 
Our contribution in this article is to identify how the velocity of sound captures the pseudogap emerging from Mott physics and resulting short-range correlations.

\section{Model and method} 

The velocity of sound $c_s$ along high symmetry directions is defined as $c_s=\sqrt{c_{ij}/\rho}$, where $\rho$ is the density, $c_{ij}=\partial^2 F/\partial u^2_{ij}$ is the elastic constant in the Voigt notation, $F$ is the free energy, and $u_{ij}$ is the strain, also in Voigt notation. 
Cuprates have tetragonal ($D_{4h}$) symmetry, and in this work we focus on the uniform compressive strain in the $xy$ plane, which is associated to the elastic constant $c_{A_{1g}}=(c_{11}+c_{22})/2$ of the $A_{1g}$ irreducible representation.  
Furthermore, we consider both electronic and lattice degrees of freedom. Hence the free energy is the sum of two terms: the electronic free energy $F_{\rm el}$ and the lattice free energy $F_{\rm latt}$, with $F_{\rm el} = F_0+F_{\rm el-latt}$, where $F_0$ is the free energy of the electrons at zero strain and $F_{\rm el-latt}$ is the free energy due to the interaction between electrons and the lattice. 
The elastic constant of the $A_{1g}$ mode, and hence its associated velocity of sound, is renormalised by the interaction between electrons and lattice as $c^*_{A_{1g}} =c_{A_{1g}} + \Delta c_{A_{1g}}$, where $\Delta c_{A_{1g}}$ is the correction to $c_{A_{1g}}$ due to the electron-lattice interaction, i.e.  physically it is the contribution to the elastic stiffness due to the electrons at the Fermi level. 

To find $\Delta c_{A_{1g}}$, we consider the following Su-Schrieffer-Heeger-Hubbard Hamiltonian~\cite{ssh:1979} on the 2D square lattice: 
\begin{align}
H = & -\sum_{\langle ij \rangle \sigma} t[a + (d_i-d_j)] (c^\dagger_{i\sigma} c_{j \sigma} + c^\dagger_{j\sigma} c_{i \sigma} ) \nonumber \\
& + U \sum_{i} n_{i\uparrow} n_{i\downarrow} - \mu \sum_{i\sigma} n_{i\sigma} .
\label{Hamiltonian}
\end{align}
Eq.~\ref{Hamiltonian} can be viewed as a {\it compressible} Hubbard model, as in Refs.~\cite{Majumdar1994, HassanSound2005}. 
Here $c^\dagger_{i\sigma}, c_{i\sigma}$ create and destroy, respectively, an electron at site $i$ of spin $\sigma$, $n_i=c^\dagger_{i\sigma} c_{i\sigma}$ is the number operator, $U$ is the onsite Coulomb interaction, and $\mu$ is the chemical potential. Contrary to the (standard) Hubbard model, the nearest neighbor hopping amplitude $t$ is modulated by the local change $(d_i-d_j)$ of the equilibrium lattice constant $a$, i.e. with zero wavevector as appropriate for ultrasound experiments. This Hamiltonian preserves electroneutrality. 
 
For the 2D compressible Hubbard model with nearest-neighbor hopping in Eq.~\ref{Hamiltonian}, the correction $\Delta c_{A_{1g}}$ to the elastic constant $c_{A_{1g}}$ due to the interaction between electrons and lattice is given by the second derivative of the electronic free energy with respect to hopping, $\partial^2 F_{\rm el}/\partial t^2$, which is proportional to $\partial \langle \epsilon \rangle /  \partial t$, where $t \langle \epsilon \rangle$ is the kinetic energy of the 2D Hubbard model (see derivation in the supplemental material~\cite{SupplementalMaterial}). This in turn determines the velocity of sound of the $A_{1g}$ mode. 
To compute $\Delta c_{A_{1g}}$, we solve the 2D Hubbard model with the cellular extension~\cite{maier, kotliarRMP, tremblayR} of dynamical mean-field theory~\cite{rmp}. We solve the cluster -here a $2\times2$ plaquette- quantum impurity problem using a continuous time quantum Monte Carlo method~\cite{millisRMP, patrickSkipList} in the hybridisation expansion of the impurity action. We focus on the normal state only.

\section{Phase diagram}

\begin{figure}[ht!]
\centering{
\includegraphics[width=1.\linewidth]{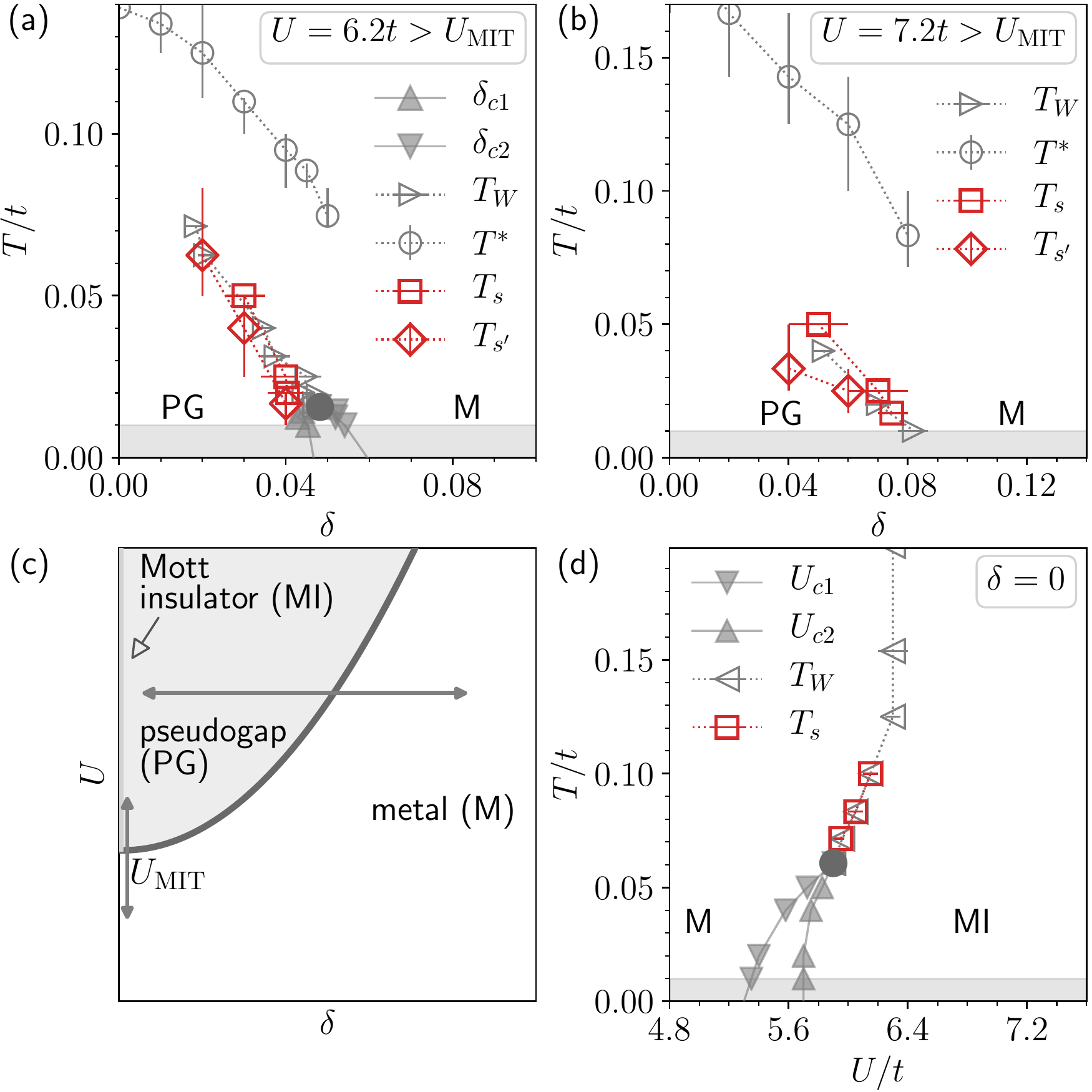}
}
\caption{(a,b) Temperature versus doping normal state phase diagram, as obtained from the CDMFT solution of the 2D Hubbard model (gray symbols). Data are obtained for $U=6.2t$ and $U=7.2t$, which set the system in the doped Mott insulator regime. 
At finite doping there is a first-order phase transition between a pseudogap (PG) and a metal (M). This transition is bounded by the spinodals $\delta_{c1}$ and $\delta_{c2}$ (filled triangles) and ends at a critical endpoint (filled dark gray circle). From the endpoint emerges the Widom line, $T_W$, here defined as the loci of the peaks in the isothermal charge compressibility $\kappa$ as a function of doping for different temperatures. The open circles denote where the spin susceptibility drops versus $T$ at fixed doping, signalling the onset temperature of the pseudogap $T^*$. 
(c) Sketch of the interaction strength $U$ versus doping normal state phase diagram of the 2D Hubbard model.
(d) Temperature versus $U$ normal state phase diagram for the half filled model ($\delta=0$). The first-order metal to Mott insulator transition is bounded by spinodals $U_{c1}$ and $U_{c2}$ (filled triangles), ending in a critical endpoint (filled circle), from which emanates the Widom line, a supercritical crossover (open triangles) defined by the loci of the inflection in the double occupancy versus $U$ at fixed $T$. 
Gray symbols are taken from Refs.~\cite{CaitlinSb, Caitlin:PRXQ2020}. Red symbols are the new results of this work and indicate the loci of the minima in the velocity of sound versus $\delta$ at fixed $T$ (squares in panels a, b) or versus $U$ at fixed $T$ (squares in panel d), or versus $T$ at fixed $\delta$ (diamonds in panels a, b). These minima are one of the main findings of this article. Shaded area at low $T$ in panels a, b, d corresponds to the region that is inaccessible because of the fermionic sign problem. 
}
\label{fig1}
\end{figure}

To understand the features of the velocity of sound of the 2D compressible Hubbard model, we briefly review the main aspects of the phase diagram of the underlying 2D Hubbard model. 
Hole-doped cuprates are doped Mott insulators. To set the system in this regime, we use a value of the interaction larger than threshold $U_{\rm MIT}$ necessary to open a Mott gap at zero doping. Figure~\ref{fig1}a shows the normal state temperature - doping phase diagram for $U=6.2t$ emerging from the CDMFT solution of the 2D Hubbard model~\cite{CaitlinSb}. At low temperature there is a first-order transition at finite doping and finite temperature between a strongly correlated metal with a pseudogap and a correlated metal. This normal state transition is actually hidden beneath a superconducting dome~\cite{sshtSC, LorenzoSC, CaitlinPNAS2021}. 
It is first-order and ends in a second-order critical endpoint (gray filled circle). From the endpoint emerge some crossover lines in the thermodynamic~\cite{ssht, sshtRHO, CaitlinOpalescence} and entanglement~\cite{Caitlin:PRXQ2020} properties with the features of the Widom line $T_W$. 
The crossover line defining the onset temperature of the pseudogap $T^*$ ends abruptly at $\delta_{p}$~\cite{sshtRHO,Alexis:2019}, as found in experiments~\cite{Collignon_Badoux_Taillefer:2017, Olivier:PRB2018}, and is a high-temperature precursor of $T_W$. The pseudogap to metal transition is a purely electronic metal to metal transition, and the two phases have the same symmetries.

To reveal the origin of this pseudogap to metal transition without symmetry breaking, it is necessary to vary the interaction $U$. Figure~\ref{fig1}b shows the $T-\delta$ phase diagram for $U=7.2t$. Upon increasing $U$, the pseudogap to metal transition moves to larger doping and lower temperature. The Widom line can be used to reveal the existence of the pseudogap to metal transition, which in Figure~\ref{fig1}b occurs at a low temperature currently inaccessible due to the fermionic sign problem~\cite{sht2}. 
By further varying the interaction $U$, one obtains~\cite{sht,sht2} the schematic $U-\delta$ normal-state phase diagram sketched in Figure~\ref{fig1}c. 
By following the horizontal arrow in the doped Mott insulator regime, the system evolves from a Mott insulator at zero doping to a pseudogap followed by a the first-order transition to a metal. The pseudogap to metal transition (thick gray line) is connected to the metal to Mott insulator transition at zero doping (vertical arrow and the resulting $U-T$ phase diagram in Figure~\ref{fig1}d). This implies~\cite{sht, sht2} that the pseudogap to metal transition originates from Mott physics and short-range correlations. Physically, this means Mott localisation plus short range correlations, arising from superexchange, form singlet bonds that open a pseudogap.

The crossing of the Widom line and its underlying second order critical line ending the first-order transition gives rise to a peak in the electronic specific heat~\cite{Giovanni:PRBcv}. 
This provides a microscopic framework that solves the apparent paradox raised by recent experiments~\cite{Michon:Cv2018} reporting a sharp peak in the low temperature normal state electronic specific heat at the doping $\delta_{p}$ where the pseudogap ends without evidence of broken symmetry states. 
Our main contribution here is to show that thermodynamic anomalies --such as the peak along the Widom line in the electronic specific heat and in the charge compressibility-- are also imprinted in the velocity of sound.  We shall show that the isothermal velocity of sound for the $A_{1g}$ mode has anomalies in the form of sharp dips versus $\delta$  upon crossing the endpoint of the pseudogap to metal transition and its associated Widom line. Red squares in Figure~\ref{fig1}a,b mark the doping levels of the dip in the velocity of sound.

\section{Velocity of sound in a doped Mott insulator}

\begin{figure}
\centering{
\includegraphics[width=1.\linewidth]{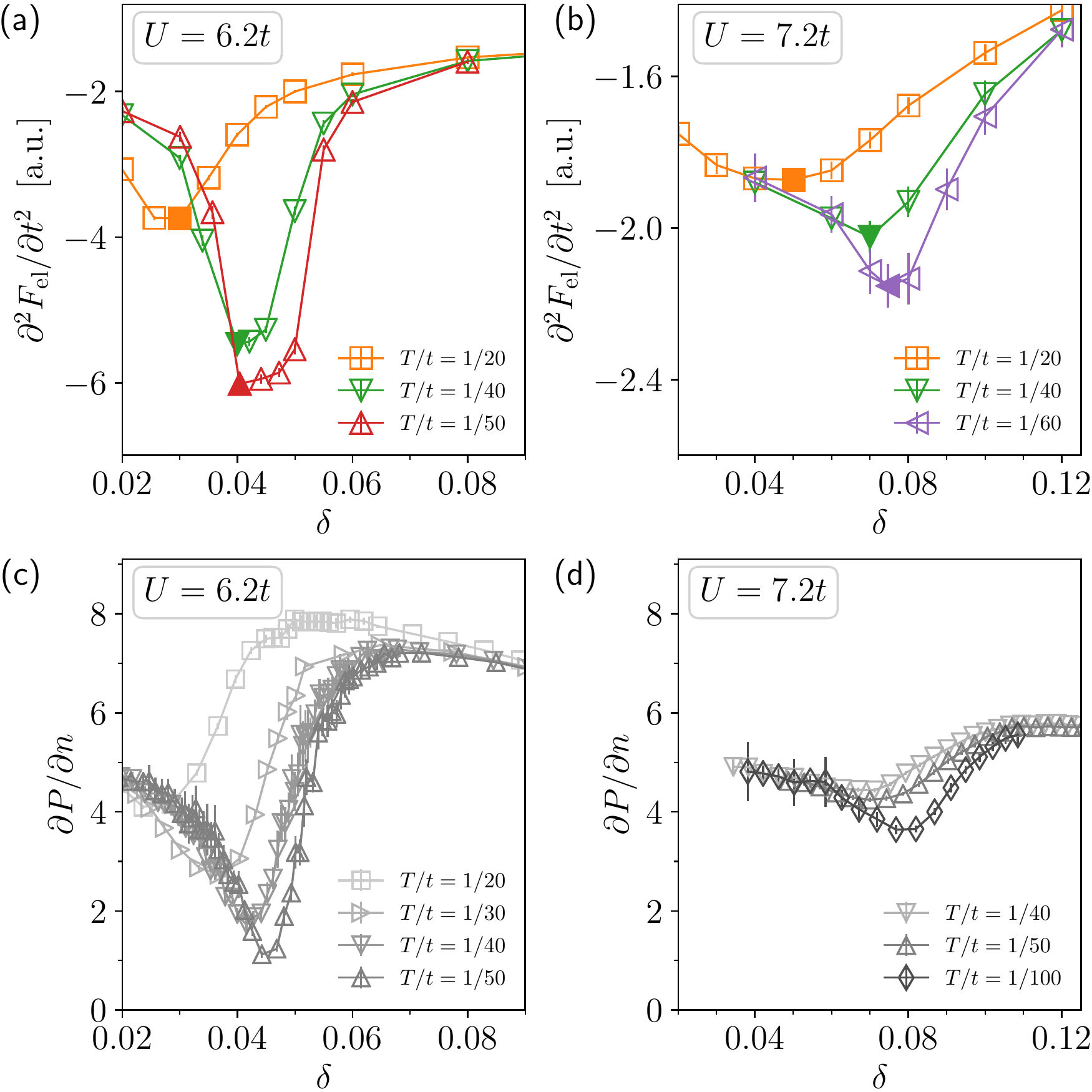}
}
\caption{Upper panels: correction $\Delta c_{A_{1g}}$ to the elastic constant $c_{A_{1g}}$ due to the electron-lattice interaction, as a function of doping and for different temperatures. This elastic constant determines the velocity of sound of the $A_{1g}$ mode. Refer to the supplemental material for the proportionality constants. Minima at finite doping are visible (filled symbols). The loci of the minima vs $\delta$ of the velocity of sound are shown by red squares ($T_s$) in Fig.~\ref{fig1}a,b. 
Lower panels: velocity of sound of the electron gas $v_s^2 \propto \partial P /\partial n$, as a function of doping and for different temperatures. Gray colors emphasise this quantity is at zero strain. 
Data are shown for $U=6.2t$ (panels a, c) and $U=7.2t$ (panels b, d). Error bars indicate the rms error.
}
\label{fig2}
\end{figure}
\begin{figure}
\centering{
\includegraphics[width=1.\linewidth]{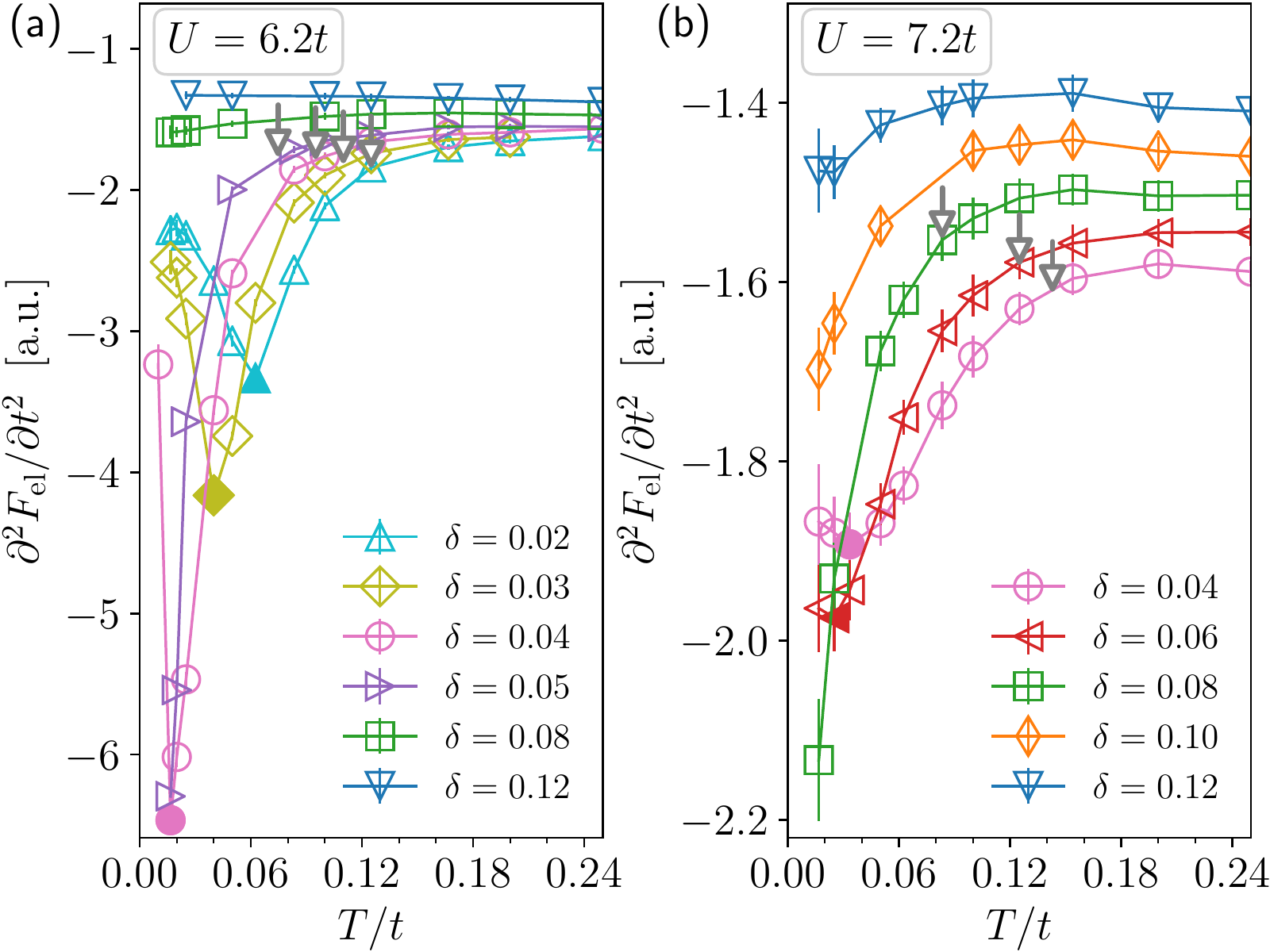}
}
\caption{Correction $\Delta c_{A_{1g}}$ to the elastic constant $c_{A_{1g}}$ due to the electron-lattice interaction, as a function of temperature for different values of doping. This quantity is proportional to the velocity of sound of the $A_{1g}$ mode. Data are obtained for $U=6.2t$ (a) and $U=7.2t$ (b). The gray arrows indicate the temperature $T^*$ at each doping level for the underlying 2D Hubbard model (gray open circles in Fig.~\ref{fig1}a,b). The curves show a dip (filled symbols) upon crossing the Widom line. The loci of the dips vs T of the velocity of sound are shown by red diamonds ($T_{s^\prime}$) in Fig.~\ref{fig1}a,b. Error bars indicate the rms error.
}
\label{fig3}
\end{figure}

Figure~\ref{fig2}a,b shows $\partial^2 F_{\rm el}/\partial t^2$ for the correction $\Delta c_{A_{1g}}$ to the elastic constant $c_{A_{1g}}$ due the electron-lattice interaction, as a function of doping for different temperatures. This term is proportional to the correction to the velocity of sound for the $A_{1g}$ mode, and has also been calculated in Ref.~\cite{HassanSound2005} in the half-filled case. 

The velocity of sound has an anomaly, in the form of a dip with a minimum (indicated by the filled symbols), as a function of doping. The dip becomes more pronounced with decreasing temperature. To understand the origin of this dip, we track the doping levels at which the dips occur for different temperatures. This result is shown in Figure~\ref{fig1}a,b on the temperature-doping phase diagram, with red squares. 
The positions of the dips define a crossover line (open red squares in Figure~\ref{fig1}a,b) emerging from the critical endpoint of the pseudogap-metal transition. 
Therefore, the anomaly in the velocity of sound for the $A_{1g}$ mode is a signature of the critical point at finite doping and finite temperature and of its associated crossover in the supercritical region. 

To further connect the anomaly in the velocity of sound with earlier results on compressibility anomaly~\cite{ssht}, we calculate the velocity of sound of the electron gas at zero strain estimated from the density-density correlation function at zero strain.  
Namely, we calculate $v_s^2 \propto \left( \partial P /\partial n \right)_T$, where $P$ is the electronic pressure at zero strain, which is associated with the electronic compressibility $\kappa=1/n^2 (\partial n / \partial \mu)$ via $\partial P /\partial n=1/(n\kappa)$~\cite{CaitlinSb, Caitlin:PRXQ2020}. Figure~\ref{fig2}c,d shows the velocity of sound of the electron gas as a function of doping for different temperatures. The curves show a dip that becomes more pronounced as the endpoint is approached. 
The presence of this minimum in the velocity of sound of the electron gas $v_s$ suggests an electronic origin of the anomaly in the velocity of sound $c_s$ for the $A_{1g}$ mode.

Next, let us discuss the onset temperature of the pseudogap, $T^*$, which is a precursory crossover of this metal to metal transition. Because of this, the velocity of sound should be continuous across $T^*$. This expectation is fully confirmed by our calculations. Figure~\ref{fig3}a,b shows the velocity of sound of the $A_{1g}$ mode as a function of temperature, for different values of doping. The vertical gray arrows indicate the temperature $T^*$ at each doping level for the underlying 2D Hubbard model (see Figure~\ref{fig1}a,b). Within our numerical uncertainty, the velocity of sound is continuous and does not show any particular features across $T^*$. 
Note that the velocity of sound shows a minimum versus $T$ (filled symbols in Figure~\ref{fig3}a,b and red diamonds in Figure~\ref{fig1}a,b) upon crossing $T_W$. The low temperature hardening signals the electrons locking into singlets due to superexchange.

\section{Velocity of sound at half filling} 

Since the pseudogap to metal transition is connected to the metal to Mott insulator transition in the interaction-doping plane (Figure~\ref{fig1}c), we shall discuss the behavior of the velocity of sound at half filling, across the metal to Mott insulator transition.

The velocity of sound of the $A_{1g}$ mode has a minimum as a function of $U/t$ (see Supplemental Material~\cite{SupplementalMaterial}). Most importantly, this minimum defines a crossover line in the supercritical region (red squares in Figure~\ref{fig1}d), which lies close to the Widom line (gray open triangles in Figure~\ref{fig1}d). This dip is a signature of the Mott critical point and its associated crossover. 
This result for the sound velocity versus $U$ at half filling confirms and extends the findings of Hassan et al.~\cite{HassanSound2005} for the 3D compressible Hubbard model solved with single-site DMFT.

\section{Discussion}  

Using the 2D compressible Hubbard model, we have calculated the isothermal velocity of sound for the $A_{1g}$ mode with CDMFT, and shown it has a sharp dip versus doping. At low temperature, this anomaly signals a second order critical endpoint at finite temperature and finite doping, which is the endpoint of the pseudogap to metal transition. This is a purely electronic transition without symmetry breaking. Upon increasing $T$ away from the endpoint, the dip becomes shallower and its locus delineates a supercritical crossover line emanating from the endpoint (red squares in Figure~\ref{fig1}a,b). 

We predict that ultrasound experiments in hole doped cuprates should observe a minimum in the velocity of sound for the $A_{1g}$ mode at the doping $\delta_{p}$ where the pseudogap ends, once superconductivity is removed by a magnetic field. Furthermore, this dip should become sharper with decreasing temperature. These predicted anomalies indicate the existence of an electronic transition without symmetry breaking, from which supercritical crossovers emerge. This transition is hidden by the superconducting phase~\cite{sshtSC, LorenzoSC, CaitlinPNAS2021, Lorenzo3band}. 
Verifying this prediction in experiments would be a worthy challenge to explore as it may help to unlock the elusive nature of the pseudogap phase.

This dip parallels the peak at $\delta_{p}$ in the electronic specific heat recently observed in the LSCO family~\cite{Michon:Cv2018, Girod:PRB2021} and successfully modeled by the 2D Hubbard model~\cite{Giovanni:PRBcv}. Given that both specific heat and velocity of sound are thermodynamic indicators of the system, our results provide additional thermodynamic constraints and show that the minimum in the velocity of sound versus $\delta$ is a likely feature of real materials. 

Our prediction for hole doped cuprates is strengthened by the following observation. The pseudogap to metal transition is connected in the $U-\delta-T$ space to the Mott transition at zero doping (Figure~\ref{fig1}c) where a minimum in the velocity of sound has already been detected in V$_2$O$_3$~\cite{Populoh2011} and in the organic superconductors~\cite{Fournier2003,HassanSound2005}, with a relative decrease of ~2\% and ~20\% respectively. 

While the low temperature crossing of $\delta_{p}$ reveals a minimum in the velocity of sound of the $A_{1g}$ mode versus doping, our calculation shows no signs of a phase transition occurring versus temperature across the onset temperature $T^*$. 
Ultrasound measurements~\cite{Shekhter2013, Frachet:NatPHys2020} find an overall hardening of the sound velocity as temperature decreases, as expected from background anharmonic-phonon effects. Since we are interested in how conduction electrons influence the sound velocity, that background must be subtracted. 
We hope that our results may guide further experimental investigations.

\begin{acknowledgments}
We are indebted to Claude Bourbonnais, Cyril Proust, and Antoine Georges for useful discussions. This work has been supported by the Canada First Research Excellence Fund. 
Simulations were performed on computers provided by the Canadian Foundation for Innovation, the Minist\`ere de l'\'Education des Loisirs et du Sport (Qu\'ebec), Calcul Qu\'ebec, and Compute Canada.
\end{acknowledgments}

\onecolumngrid
\clearpage
\setcounter{figure}{0}
\setcounter{section}{0}
\setcounter{equation}{0}
\makeatletter 
\renewcommand{\thefigure}{S\@arabic\c@figure} 

\begin{center}

{\bf Supplemental Material:} \\
\vspace{0.05cm} 

{\bf Prediction of anomalies in the velocity of sound for the pseudogap of hole-doped cuprates} \\
\vspace{0.05cm}

{C. Walsh, M. Charlebois, P. S\'emon, G. Sordi, A.-M. S. Tremblay}

\end{center}

In Section~\ref{Sec:Derivation} we derive the result for the correction to the velocity of sound $c_s$ of the $A_{1g}$ mode used in the main text. In Section~\ref{Sec:Half-filling} we provide additional data for the velocity of sound at half filling.

\section{Derivation of the correction to the velocity of sound}
\label{Sec:Derivation}

\subsection{Definitions}

The velocity of sound $c_s$ along high symmetry directions is $c_s=\sqrt{c_{ij}/\rho}$, where $\rho$ is the density and $c_{ij}$ is the elastic constant in the Voigt notation~\cite{LandauElasticityBOOK}. Note that $c_{ij}$ has the units of pressure and is defined as the second derivative of the free energy $F$ with respect to the strain $u_{ij}$, $c_{ij}=\partial^2 F/\partial u_{ij}^2$, with 
\begin{align}
u_{ij} & = \frac{1}{2} \left( \frac{\partial u_i}{\partial r_j} + \frac{\partial u_j}{\partial r_i} \right) ,
\end{align}
where $u_i({\bf r})$ is the displacement along the direction $r_i$, with $(r_1,r_2,r_3)=(x,y,z)$. Note that the strain $u_{ij}$ is dimensionless, and thus $F_{\rm latt}$ has the units of $c_{ij}$, i.e. it is a free energy density.

We want to capture the physics of the cuprates, which have tetragonal crystal structure. Hence we consider the point group $D_{4h}$. The resulting lattice free energy $F_{\rm latt}$ is~\cite{Benhabib:NatPhys2020} 
\begin{align}
F_{\rm latt} & = \frac{1}{2} c_{11} (u_{xx}^2 + u_{yy}^2) + c_{12} u_{xx}u_{yy} + 2c_{66} u_{xy}^2 + \frac{1}{2} c_{33} u_{zz}^2 + c_{13} (u_{xx} + u_{yy}) u_{zz} ,  
\label{F1a}
\end{align}
with standard Voigt notation for the elastic constants. Furthermore, cuprates are layered materials of 2D planes stacked along the $z$ axis. We shall calculate the longitudinal velocity of sound within the $xy$ plane. Hence we consider only the strain in the $xy$ plane, and neglect the strain in the $z$ direction. Hence we are left to consider the lattice free energy 
\begin{align}
F_{\rm latt} & = \frac{1}{2} c_{11} (u_{xx}^2 + u_{yy}^2) + c_{12}u_{xx}u_{yy}.  
\label{F1b}
\end{align}

Furthermore, we shall consider only uniform compressive strain. Hence, it is useful to write $F_{\rm latt}$ using the irreducible representations $A_{1g}$ and $B_{1g}$ of the $D_{4h}$ group. The uniform compression is associated with the $A_{1g}$ mode. The $A_{1g}$ and $B_{1g}$ components of the strain can be written as 
\begin{align}
u_{A_{1g}} & = u_{xx} + u_{yy}  \label{sm:uA1g} \\
u_{B_{1g}} & = u_{xx} - u_{yy} .  \label{sm:uB1g}
\end{align}
Hence Eq.~\ref{F1b} becomes: 
\begin{align}
F_{\rm latt} & = \frac{1}{2} \left( c_{A_{1g}} u_{A_{1g}}^2 + c_{B_{1g}} u_{B_{1g}}^2 \right),
\end{align}
where the elastic constants for the $A_{1g}$ and $B_{1g}$ modes are
\begin{align}
c_{A_{1g}} & = \frac{1}{2} (c_{11} + c_{12}) \\
c_{B_{1g}} & = \frac{1}{2} (c_{11} - c_{12}) .
\end{align}
Therefore the longitudinal velocity of sound that we shall calculate is 
\begin{align}
c_s & = \sqrt{\frac{c_{A_{1g}}}{\rho}} .
\end{align}

Up until now we have only considered lattice degrees of freedom. We want to describe both lattice and electronic degrees of freedom, i.e. we want to consider 
\begin{align}
F & = F_{\rm latt} + F_{\rm el} 
\end{align}
with
\begin{align}
F_{\rm el} & = F_{\rm 0} + F_{\rm el-latt} , 
\end{align}
where $F_{\rm 0}$ is the free energy of the electrons at zero strain. As a result, the velocity of sound $c_s$ is renormalised by the interaction with the electrons. Hence we can write: 
\begin{align}
c_s & = \sqrt{\frac{c^*_{A_{1g}}}{\rho}} ,
\end{align}
where 
\begin{align}
c^*_{A_{1g}} = c_{A_{1g}} + \Delta c_{A_{1g}} .
\end{align}
The correction $\Delta c_{A_{1g}}$ is coming from the interaction with the electrons. Our goal is thus is to find an estimate for $\Delta c_{A_{1g}}$.

\subsection{2D compressible Hubbard model}

To find $\Delta c_{A_{1g}}$, we consider the Su-Schrieffer-Heeger (SSH) Hubbard model~\cite{ssh:1979} on a 2D square lattice. To simplify the derivation, we first consider a strain applied in only one direction. Since we consider only nearest neighbor hopping, $x$ and $y$ directions are independent. Potential energy is unaffected, so we need only consider how strain modifies the kinetic energy in the direction of the lattice vector where it is applied, hence we have 
\begin{align}
-\sum_{\langle ij \rangle \sigma} t[a + (d_i-d_j)] (c^\dagger_{i\sigma} c_{j \sigma} + c^\dagger_{j\sigma} c_{i \sigma} ) ,   
\label{sm:2DcompressibleHM}
\end{align}
where $i$ and $j$ now refer to lattice positions, not cartesian directions as before. We take $d_i$ as the displacement from equilibrium of the atom at position $i$. This model can be also referred to as a {\it compressible} Hubbard model, as in Refs.~\cite{Majumdar1994, HassanSound2005}. Here $c^\dagger_{i\sigma}$ and $c_{i\sigma}$ operators create and destroy an electron at site $i$ of spin $\sigma$. The equilibrium distance between atoms is $a$. 
Contrary to the standard Hubbard model, the nearest neighbor hopping amplitude $t$ in Eq.~\ref{sm:2DcompressibleHM} is not constant, but is modulated by the local change $(d_i-d_j)$ of the equilibrium lattice constant $a$.

A Taylor expansion of the hopping amplitude $t$ about $(d_i-d_j)=0$ gives:
\begin{align}
t[a + (d_i-d_j)] \approx t_{ij} + \frac{\partial t}{\partial a} (d_i -d_j) ,
\end{align}
where the equilibrium value $t(a)$ is denoted by $t_{ij}$, which is equal to $t$ for nearest neighbor hopping and to zero otherwise. Therefore the hopping term $K_{\rm el}$ that is modified by $d_i -d_j$ can be written as: 
\begin{align}
K_{\rm el} & = K_{0} + K_{\rm el-latt} \\
& = -\sum_{\langle ij \rangle \sigma } t_{ij} (c_{i \sigma}^\dagger c_{j\sigma} +c_{j \sigma}^\dagger c_{i\sigma} ) - \sum_{\langle ij \rangle \sigma } \frac{\partial t}{\partial a} (d_i -d_j) (c_{i \sigma}^\dagger c_{j\sigma} +c_{j \sigma}^\dagger c_{i\sigma} ) ,  
\label{K_tot}
\end{align}
where $K_{0}$ is the familiar hopping term in the standard Hubbard model at zero strain and $K_{\rm el-latt}$ is the modulated hopping term as in a SSH-like model.

It is convenient to take the origin of each bond in the middle of two sites. Along the $x$ direction then, we rewrite the change in local interatomic distance in terms of the strain, using the notation of the previous subsection in the long wave-length limit 
\begin{align}
d_{i} - d_{i+1} = - \left . \frac{\partial u_x({\bf r})}{\partial x} \right|_{{\bf r}_k} a ,
\end{align}
where ${\bf r}_k = ({\bf r}_{i+1} - {\bf r}_i)/2$, and where index $k$ labels bonds. We rewrite the second-quantized operators in this basis and define $\epsilon_x ({\bf r}_k)$ as follows 
\begin{align}
c_{i}^\dagger c_{i+1} = c^\dagger_{{\bf r}_k -a/2} c_{{\bf r}_k +a/2} = \epsilon_x ({\bf r}_k) .
\end{align}
because stretched bonds are in the $x$ direction.

Then the electron-lattice interaction $K_{\rm el-latt}$ in Eq.~\ref{K_tot} along the $x$ direction becomes 
\begin{align}
\sum_{x_k} \left( - a \frac{\partial t}{\partial a} \right) u_{xx} ({\bf r}_k) \left( c^\dagger_{{\bf r}_k -a/2} c_{{\bf r}_k +a/2} +h.c.\right) 
& = g \sum_{x_k} u_{xx} ({\bf r}_k) \left( \epsilon_x ({\bf r}_k) + h.c. \right),  
\label{K_el-latt2}
\end{align}
with the definition
\begin{align}
g & = - a \frac{\partial t}{\partial a} .
\end{align}
The result is similar in the $y$ direction. Note that $g$ has units of energy and that $g>0$ is satisfied because $t$ decreases when $a$ increases. 

Next we move to reciprocal space assuming a single wave-vector. And since we consider uniform compressive strain we will eventually take the $q\rightarrow 0$ limit. Taking the Fourier transform of the strain gives: 
\begin{align}
u_{xx} ({\bf r}_k) & = u_{xx} ({\bf q}) \left( e^{i{\bf q} \cdot {\bf r}} + e^{-i{\bf q} \cdot {\bf r}} \right)/2 ,
\end{align}
where ${\bf q}$ is in the $x$ direction and $q \rightarrow 0$. Then the interaction Eq.~\ref{K_el-latt2} becomes (since $u_{xx}(q) = u_{xx}(-q)$): 
\begin{align}
g \frac{u_{xx}(q)}{2} \left( \epsilon^-_x (-q) +\epsilon^+_x(-q) \right) + g \frac{u_{xx}(-q)}{2} \left( \epsilon^-_x (q) +\epsilon^+_x(q) \right) ,
\end{align}
with a similar result in the $y$ direction. Here $\epsilon^+$ and $\epsilon^-$ denote the two terms entering the kinetic energy. Both $u_{xx}(q)$ and $\epsilon^\pm_x$ are dimensionless.

Bringing together the terms along the $x$ and $y$ directions, we obtain: 
\begin{align}
K_{\rm el-latt} & = g \left[ \epsilon_x(-q) u_{xx}(q)/2 + \epsilon_x(q) u_{xx}(-q)/2 \right] + g \left[ \epsilon_y(-q) u_{yy}(q)/2 + \epsilon_x(q) u_{yy}(-q)/2 \right] ,
\end{align}
where we defined:
\begin{align}
\epsilon_x(q) = \epsilon^-_x(q) + \epsilon^+_x(q) \\
\epsilon_y(q) = \epsilon^-_y(q) + \epsilon^+_y(q).
\end{align}

Our aim is to trace over the electronic degrees of freedom to find out how $F_{\rm latt}$ is modified. For completeness then, let us also write $F_{\rm latt}$ in reciprocal space. For clarity, let us focus only on $u_{xx}$ with a single wave-vector as above. We find 
\begin{align}
F_{\rm latt} & = \int \frac{1}{2} c_{11} u_{xx} ({\bf r}) u_{xx} ({\bf r}) \, d{\bf r}= \frac{1}{2} c_{11} \frac{u_{xx}(q)u_{xx}(-q)}{2} .  
\label{F_function_of_uxx}
\end{align}

\subsection{Correction to the elastic free energy from coupling to the conduction electrons}

We take the following steps, valid only for a Hubbard model with nearest neighbor hopping. Otherwise, the operator for the stress tensor is more complicated. First we find the correction to the elastic free-energy to second-order in $g$ for a simple strain $u_{xx}$. We obtain a zero-frequency Matsubara response function as the correction, a result identical to what we would have obtained from a phonon self-energy-calculation for this model. Second we trivially extend the result to a $A_{1g}$ strain. Finally, we rewrite the result for the correction to the elastic constant $c_{A_{1g}}$ in terms of a derivative of the kinetic energy.

\subsubsection{Perturbative result for the free energy}
\label{Perturbative}

The contribution to the elastic stiffness due to the conduction electrons is obtained by tracing over them. In other words, in this section we are focusing on the contributions to the partition function that comes from the electronic Hamiltonian and from the interaction of the electrons with the lattice. We assume that the elastic free energy without the conduction electrons is already known.

As before, we can separate the strains along the $x$ and $y$ direction. Let us focus on the $x$ direction. We take the strain $u_{xx}$ as a {\it classical} variable that commutes with the electronic Hamiltonian. We can also now take the $q\rightarrow 0$ limit without worrying because $\epsilon _{x}$ is not conserved, so the $q\rightarrow 0$ limit and the zero-frequency limit can be taken in any order. Therefore, in the interaction representation, the partition function of the electrons including the contributions of the strain is: 
\begin{align}
Z & = {\rm Tr}_{\rm el} \left[ e^{-\beta H_{\rm el}} T_{\tau} e^{-\int_0^\beta d\tau g \left[ \epsilon_x(0, \tau) u_{xx}(0) \right] } \right] ,
\label{eq:Z_u}
\end{align}
where $T_{\tau}$ is the imaginary-time ordering operator, $H_{\rm el}$ is the electronic Hamiltonian $\left( -\mu N\right) $ that includes the Hubbard term and the operator $\epsilon_x(0,\tau)$ in imaginary time is given by the usual Heisenberg imaginary-time evolution 
\begin{align}
\epsilon_x(0,\tau) & = e^{H_{\rm el} \tau} \epsilon_x(0) e^{-H_{\rm el} \tau} . 
\end{align}

\begin{figure}
\includegraphics[width=0.6\linewidth]{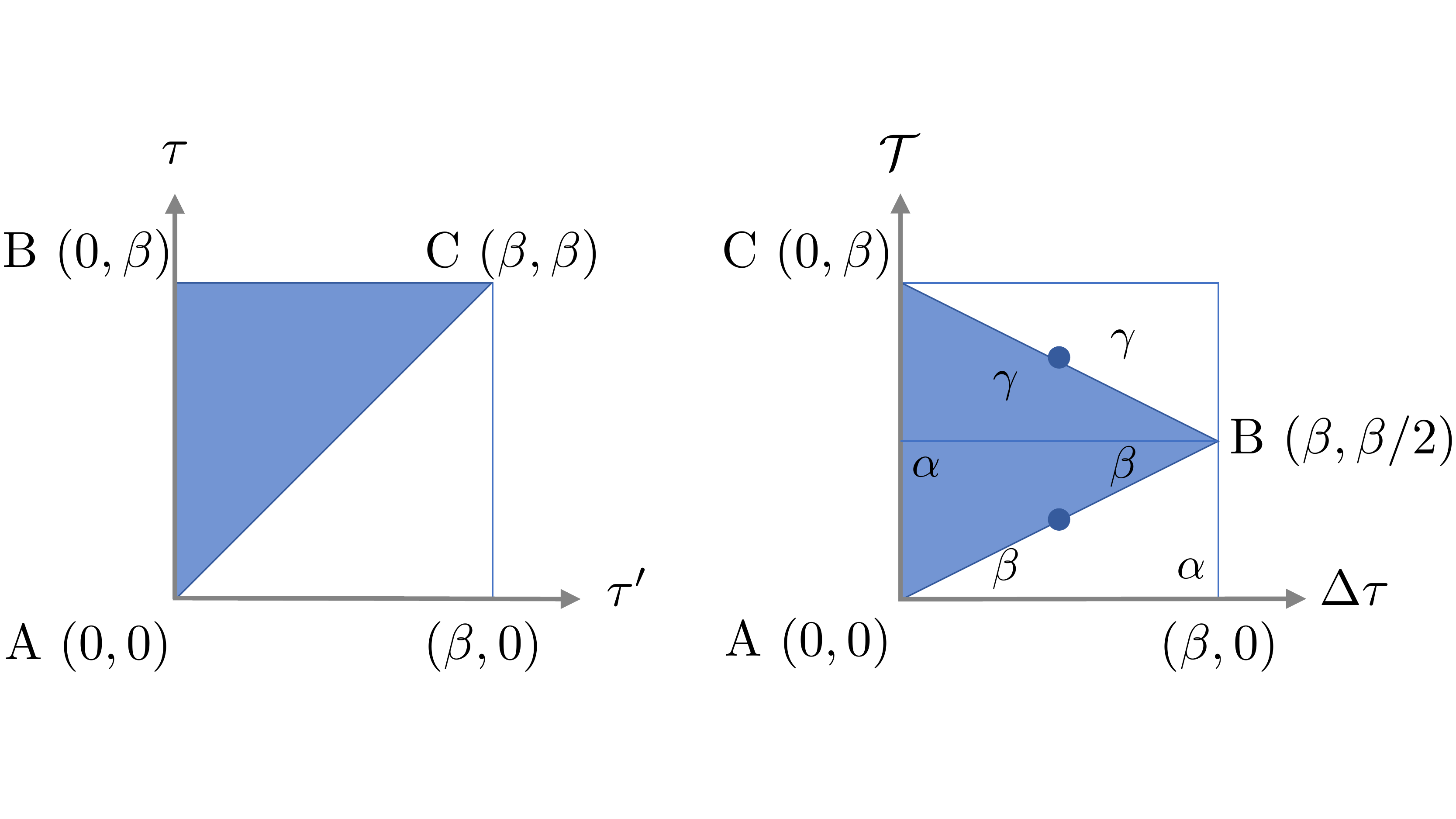}
\caption{Change of the integration domain for the time ordered product, on the left, with the change of variables ${\cal T}=(\tau+\tau')/2$; $\Delta\tau=\tau-\tau'$ on the right. The $(\tau,\tau')$ coordinates, abbreviated A, B, C on the left keep the same labels in the $(\Delta\tau, {\cal T})$ coordinates on the right. The greek letters indicate examples of points where the function has the same value given that the integrand is independent of ${\cal T}$ and that Eq.~\ref{eq:reflection} implies that $\langle \epsilon_x(\Delta\tau -\beta/2)\epsilon_x(0) \rangle = \langle \epsilon_x (\beta/2-\Delta\tau) \epsilon_x(0) \rangle$. The points on the lower two triangles (blue and transparent) are related by inversion symmetry about $(\Delta\tau/2,\beta/4)$ and those on the upper two triangles by inversion symmetry about $(\Delta\tau/2,3\beta/4)$. Blue dots mark the latter points.
}
\label{fig:integration}
\end{figure}

Expanding the exponential to second order, we obtain: 
\begin{align}
Z & = Z_{\rm 0} \left\langle 1 - \int_0^\beta d\tau g \left[ \epsilon_x(0, \tau) \right] u_{xx}(0)  +\frac{1}{2} \int_0^\beta d\tau \int_0^\beta d\tau^{\prime} g^2 T_{\tau} \left[\epsilon_x(0,\tau) \epsilon_x(0,\tau^{\prime}) \right] u_{xx}(0)u_{xx}(0)  \right\rangle .
\end{align}
Here the brackets refer to a thermal average with the electronic Hamiltonian at zero strain. The partition function at zero strain is denoted $Z_{\rm0}$ and the corresponding free energy $F_{\rm 0}$. For conciseness, we drop from now on the label that indicates that the variables are evaluated at $q=0$. Finally then, the contribution to the free energy from the conduction electrons is given by 
\begin{align}
F_{\rm el} &= -\frac{1}{\beta} \ln Z = F_{\rm 0} -\frac{1}{\beta} \ln \left[ \langle 1 \rangle - g\langle X\rangle +g^2\langle Y \rangle \right] ,
\end{align}
where 
\begin{align}
X & = \int_0^\beta  d\tau \epsilon_x(\tau )u_{xx}
\end{align}
is first order in $u_{xx}$ and 
\begin{align}
Y & =\frac{1}{2}\int_0^\beta d\tau \int_0^\beta d\tau^{\prime} \, T_{\tau} \left[ \epsilon_x(\tau) \epsilon_x(\tau^{\prime}) \right] u_{xx} u_{xx}
\end{align}
is second order in $u_{xx}$. Expanding the natural logarithm to second order in $gu_{xx}$ we thus obtain: 
\begin{align}
F_{\rm el} & = F_{\rm 0} -\frac{1}{\beta} \left[ g\langle -X\rangle +g^2\langle Y\rangle -\frac{g^2}{2}\langle X\rangle^2 \right] .
\end{align}
The first term in the square brackets is a shift of the zero of the strain. Substituting the above definitions of $X$ and $Y$ in the last equation, we are left with 
\begin{align}
F_{\rm el} & = F_{\rm 0} -\frac{g^2}{\beta} \left[ \int_0^\beta d\tau \int_0^\beta d\tau^{\prime}  \langle T_{\tau}  \epsilon_x (\tau) \epsilon_x (\tau^{\prime}) \rangle -\left[ \int_0^\beta d\tau \langle {\epsilon_x(\tau)}\rangle \right]^2 \right] \frac{u_{xx} u_{xx} }{2} .
\label{Free_energy}
\end{align}
The cyclic property of the trace, equivalently imaginary-time translation invariance, leads to $\langle {\epsilon_x (\tau)}\rangle=\langle {\epsilon_x}\rangle$ independent of $\tau$. In the first term, we need to take into account the definition of the time-ordering operator, which leads to
\begin{align} 
\int_0^\beta d\tau \int_0^\beta d\tau^{\prime} \, \langle T_{\tau}{\epsilon_{x}(\tau)\epsilon_{x}(\tau ^{\prime})} \rangle 
& = 2\int_0^\beta d\tau \int_0^\tau d\tau^\prime \langle \epsilon_{x}(\tau) \epsilon_{x}(\tau^\prime) \rangle .
\label{time_order}
\end{align}
Changing variables to 
\begin{align}
{\cal T} & = (\tau+\tau^\prime)/2 \, ; \quad \Delta\tau  = \tau - \tau^\prime
\end{align}
the Jacobian is unity and the integral becomes
\begin{align}
2 \int_0^\beta d\tau \int_\tau^\beta d\tau^{\prime} \langle \epsilon_{x}(\tau^\prime) \epsilon_x(\tau) \rangle 
& = 2 \left[ \int_0^{\beta/2} d{\cal T} \int_0^{l(\cal T) } d\Delta\tau \langle \epsilon_{x}(\Delta\tau) \epsilon_{x}(0) \rangle + \int_{\beta/2}^\beta d{\cal T} \int_0^{u(\cal T)} d\Delta\tau \langle \epsilon_{x}(\Delta\tau) \epsilon_{x}(0) \rangle \right] \\
& = \beta \int_0^\beta d\Delta\tau \langle \epsilon_{x}(\Delta\tau) \epsilon_{x}(0) \rangle .
\label{time_order_new_variables}
\end{align}
The last equality can be proven as follows. The domain of integration in the ${\cal T}, \Delta\tau$ plane on the right-hand side of the first equality is displayed to the right of Fig.~\ref{fig:integration}. The integral over $\Delta\tau$ for a given ${\cal T}$ is over the length of the segment starting at $0$ and ending at the lower $l({\cal T})$ or upper $u({\cal T})$ boundaries of the triangle on the right of the figure. However, as we show momentarily, the equality 
\begin{align}
\langle {\epsilon_x(\Delta\tau)\epsilon_x(0)} \rangle & = \langle \epsilon_x(\beta-\Delta\tau)\epsilon_x(0) \rangle
\label{eq:reflection}
\end{align}
holds, so that twice the integral over the triangular region is equal to the integral over the square (see caption of Fig.~\ref{fig:integration}), leading to the desired result Eq.~\ref{time_order_new_variables} since the integral over $\Delta\tau$, with $\Delta\tau$ always from $0$ to $\beta$, is independent of ${\cal T}$ and the integral over ${\cal T}$ gives $\beta$. To prove the last equality, Eq.~\ref{eq:reflection}, it suffices to use the cyclic property of the trace, as in the equation below, where $\psi_x$ has been used as a proxy for $\epsilon_x$ to ease the reading of the proof:
\begin{align}
{\rm Tr} \left[ e^{-\beta H} e^{H\Delta\tau} \epsilon_x e^{-H\Delta\tau} \psi_x \right]
& = {\rm Tr} \left[ e^{-H\Delta\tau} \psi_x e^{-\beta H} e^{H\Delta\tau} \epsilon_x \right] \\
&={\rm Tr} \left[ \left( e^{-\beta H} e^{\beta H} \right) e^{-H\Delta\tau} \psi_x e^{-\beta H} e^{H\Delta\tau} \epsilon_x \right] .
\end{align}

Substituting the result of the integral Eq.~\ref{time_order_new_variables} in our previous expression for the free-energy Eq.~\ref{Free_energy}, the final expression takes the form 
\begin{align}
F_{\rm el} & = F_{\rm 0} -{g^2} \left[ \int_0^\beta d\tau \, \langle {\epsilon_x (\tau) \epsilon_x (0)} \rangle -\beta \langle {\epsilon_x}\rangle^2 \right] \frac{u_{xx} u_{xx} }{2} .
\label{F3}
\end{align}
The renormalization of the elastic constant that one can read from this equation is exactly what one would have obtained from the polarization operator for the self-energy of phonons in the nearest neighbor SSH-Hubbard model for $q\rightarrow 0$. The zero-frequency Matsubara frequency appears in the answer, as we expect for thermodynamic quantities. In general this is the quantity we should compute. It is quite complicated to do this calculation rigorously including vertex corrections. We are going to find an expression that is more convenient to obtain numerically and that contains vertex corrections. But first, let us find the expression for a $A_{1g}$ strain.

\subsubsection{Response to the $A_{1g}$ mode}

By applying a uniform strain in both the $x$ and $y$ directions, we can find the response to a $A_{1g}$ mode. First, we invert Eqs.~\ref{sm:uA1g}, \ref{sm:uB1g} and find 
\begin{align}
u_{xx} & =(u_{A_{1g}} + u_{B_{1g}})/2 \\
u_{yy} & =(u_{A_{1g}} - u_{B_{1g}})/2 .
\end{align}
Rewriting the coupling of the strain in both directions in terms of irreducible representations, we find 
\begin{align}
u_{xx}\epsilon_x+u_{yy} \epsilon_y & = \frac{1}{2} u_{A_{1g}} (\epsilon_x +\epsilon_y) + \frac{1}{2} u_{B_{1g}} (\epsilon_x -\epsilon_y ).
\end{align}
If we have only $u_{A_{1g}}$, we couple to the total kinetic energy at $q\rightarrow 0$. We can redo the above derivation to compute the correction to the free energy for a $A_{1g}$ mode. The steps are identical. Hence we find 
\begin{align}
F_{\rm el} & = F_{\rm 0} - \frac{g^2}{4} \left[ \int_0^\beta d\tau \, \langle {\epsilon (\tau) \epsilon(0)}\rangle   -\beta \langle \epsilon \rangle^2 \right] \frac{1}{2} u_{A_{1g}}^2 ,
\label{F4}
\end{align}
which is valid only for nearest neighbors, as we have already mentioned, i.e. for 
\begin{align}
\epsilon & =\epsilon_x(q=0) +\epsilon_y(q=0) \\
& =2\sum_{k\sigma}(\cos k_x a+\cos k_y a) \, c_{k\sigma}^\dagger c_{k\sigma}.
\end{align}

\subsubsection{Correction to the velocity of sound from a derivative of the kinetic energy}

Finally, we show that $\int_{0}^{\beta }d\tau \langle T_\tau \epsilon(\tau) \epsilon(0) \rangle -\beta \langle \epsilon \rangle^2$ in Eq.~\ref{F4} can be evaluated from a derivative of the expectation value for the kinetic energy of the purely electronic Hamiltonian. The expectation value of $\epsilon $ at zero strain is 
\begin{align}
\langle \epsilon \rangle & = \frac{{\rm Tr}_{\rm el} \left[ \epsilon \, e^{-\beta \left[ -t\epsilon + V \right] } \right] }{ {\rm Tr}_{\rm el} \left[ e^{-\beta \left[ -t\epsilon +V \right] }\right] } ,
\end{align}
where $V$ contains the Hubbard interaction and the chemical potential contribution. This can be computed from 
\begin{align}
\langle \epsilon \rangle & = \frac{1}{\beta} \frac{\partial}{\partial t} \ln Z .
\end{align}
even if the commutator $[\epsilon , H_{\rm el}]$ does not vanish, because the power series of the exponential can be differentiated term by term and re-exponentiated at the end using the cyclic property of the trace. If we take a second derivative, the cyclic property cannot be used to rearrange the expansion of the exponential in a useful manner. 

The correct way to approach this problem is to consider the variable $t+dt$ so that $-\epsilon dt$ is now a perturbation that does not commute with $H_{\rm el}$. We then need to use the interaction representation so that the partition function is given by 
\begin{align}
Z & = {\rm Tr}_{\rm el} \left[ e^{-\beta H_{\rm el}} T_{\tau} e^{\int_0^\beta d\tau \, \left( \epsilon (\tau) dt \right) } \right] .
\end{align}
Already here, it is clear that the structure of the equation is identical to Eq.~\ref{eq:Z_u} obtained in terms of $u_{xx}$, namely $-gu_{xx}$ is here replaced by $dt$. So we already know the answer. 

Nevertheless, let us give an alternate derivation. Under a time-ordered product, derivatives act as derivatives of real numbers. Hence, to first order
\begin{align}
\frac{1}{\beta }\frac{\partial}{\partial (dt)} \ln Z 
& = \frac{1}{\beta Z} {\rm Tr}_{\rm el} \left[ e^{-\beta H_{\rm el}} T_{\tau} e^{\int_0^\beta d\tau \epsilon(\tau) dt} \left( \int_0^\beta d\tau \epsilon(\tau) \right) \right] .  
\label{PermierOrdre}
\end{align}
The above equation evaluated at zero strain, $dt=0$, leads to 
\begin{align}
\frac{1}{\beta} \frac{\partial}{\partial t} \ln Z 
& = \frac{1}{Z} {\rm Tr}_{\rm el} \left[ e^{-\beta H_{\rm el}} \epsilon \right] = \langle \epsilon \rangle ,
\end{align}
where we have used the cyclic property of the trace, which is here equivalent to invariance under imaginary-time translation. The chain rule tells us that derivatives with respect to $dt$ are identical to derivatives with respect to $t$. 

The second-order term is obtained by expanding $\ln Z$ to second order without forgetting that after the first order derivative, the partition function $Z$ appearing in the denominator must also be expanded. We find 
\begin{align}
\frac{1}{\beta^2} \frac{\partial^2}{\partial (dt)^2} \ln Z 
& = \frac{1}{\beta^2 Z} {\rm Tr}_{\rm el} \left[ e^{-\beta H_{\rm el}} T_{\tau} e^{\int_0^\beta d\tau \, \epsilon (\tau) dt} \,  \left(\int_0^\beta d\tau \, \epsilon (\tau) \right) \left( \int_0^\beta d\tau^\prime \, \epsilon (\tau^\prime)\right) \right] -\langle \epsilon \rangle^2 ,
\end{align}
where the last term comes from the expansion of $Z$ in the denominator of the first-order result. Evaluating at zero strain, $dt=0$, and using the definition of the time-ordered product along with invariance under imaginary-time translation as we did from Eqs.~\ref{time_order} to \ref{time_order_new_variables}, we are left with
\begin{align}
\frac{1}{\beta^2} \frac{\partial^2}{\partial t^2} \ln Z 
& = \frac{1}{\beta} \frac{\partial}{\partial t} \langle \epsilon \rangle = \frac{1}{\beta Z} {\rm Tr}_{\rm el} \left[ e^{-\beta H_{\rm el}} \int_0^\beta \epsilon (\tau) \epsilon (0) d\tau \right] - \langle \epsilon \rangle^2 \label{eq:d2lnZ} \\ 
& = \frac{1}{\beta} \int_0^\beta d\tau \, \langle {\epsilon (\tau) \epsilon (0)} \rangle  - \langle \epsilon \rangle^2 .
\end{align}

Since the free energy $F_{\rm el} = F_{\rm 0} + F_{\rm el-latt}$ is defined by $F_{\rm el}=\frac{1}{\beta} \ln Z$, we can substitute this last result in our expression for the free energy Eq.~\ref{F4} to find
\begin{align}
F_{\rm el} & = F_{\rm 0} -\frac{g^2}{4} \frac{\partial \langle \epsilon \rangle }{\partial t} \frac{1}{2} u^2_{A_{1g}} .
\end{align}
Therefore, the correction $\Delta c_{A_{1g}}$ to the elastic constant $c_{A_{1g}}$ due to the electron-lattice interaction is 
\begin{align}
\Delta c_{A_{1g}} & = \frac{\partial^2 F_{\rm el}}{\partial u^2_{A_{1g}}} \\
& =\left. -\frac{g^2}{4} \frac{\partial \langle \epsilon \rangle }{\partial t}\right\vert _{U,T,\mu} , 
\label{eq:Deltac}
\end{align}
which, using $ F_{\rm el} = -\ln Z/\beta$ and Eq.~\ref{eq:d2lnZ} may be rewritten as 
\begin{align}
\Delta c_{A_{1g}} & =  \frac{g^2}{4} \frac{\partial^2 F_{\rm el}}{\partial t^2} .
\end{align}
This has units of energy density as required. In Eq.~\ref{eq:Deltac} we have specified explicitly that the partial derivative is at constant $U,T,\mu$.

For the numerical results in the main text, we compute the derivative $\partial \langle \epsilon \rangle / \partial t$ in Eq.~\ref{eq:Deltac} using centered differences with respect to $t$, with a mesh of $\Delta t = 0.005$. To obtain the kinetic energy $t\langle \epsilon \rangle$ we use the method described in Ref.~\cite{LorenzoSC}.

\section{Extended data for the velocity of sound at half filling}
\label{Sec:Half-filling}

\begin{figure}[h!]
\centering{
\includegraphics[width=0.58\linewidth]{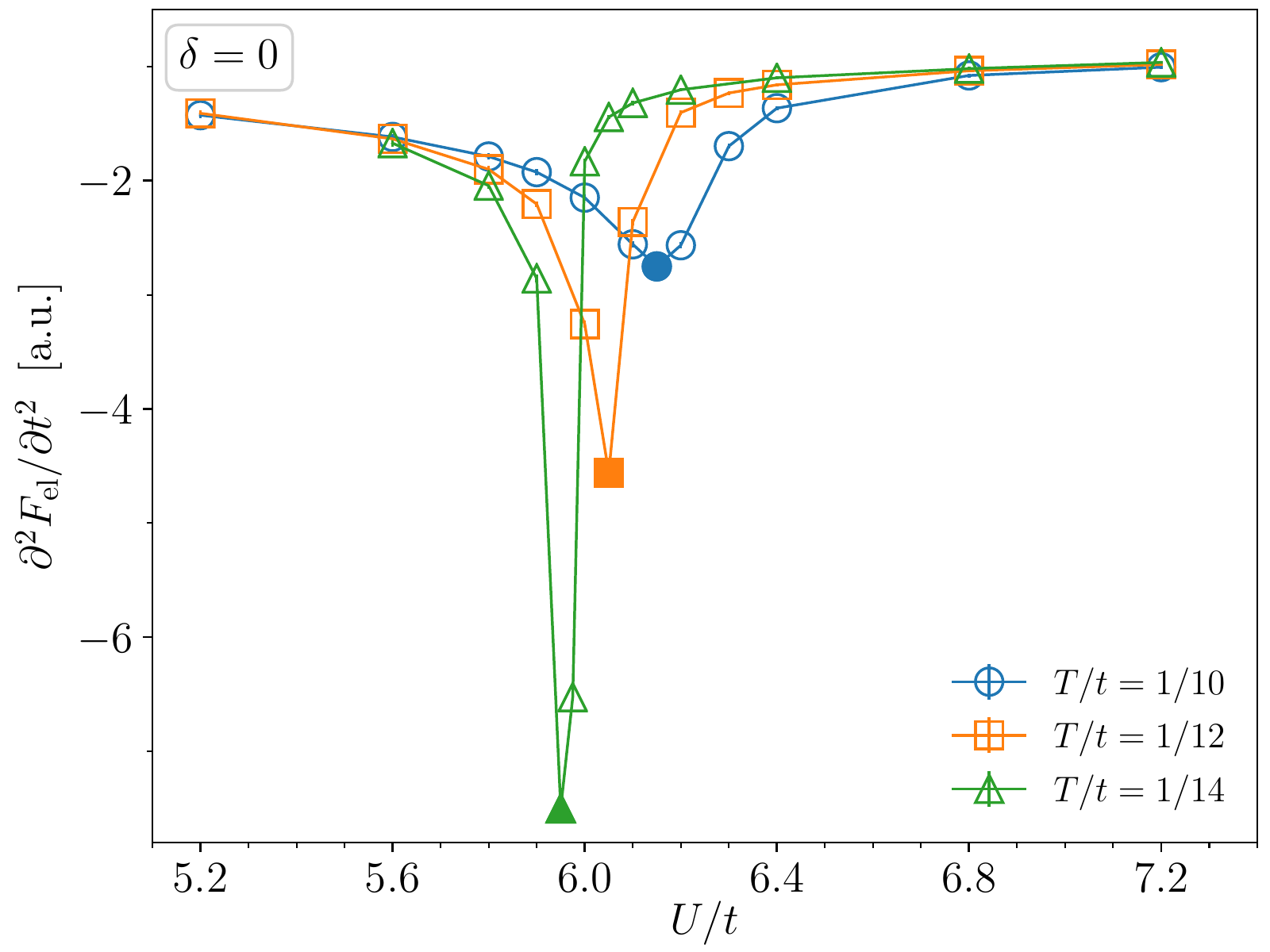} 
}
\caption{Correction $\Delta c_{A_{1g}}$ to the elastic constant $c_{A_{1g}}$ due to the electron-lattice interaction, at half filling ($\delta=0$) as a function of $U/t$ and for different temperatures. A minimum at finite $U$ and finite $T$ is visible (filled symbols). The loci of the minima of the velocity of sound are shown by red squares in the $T-U$ phase diagram of Fig.~1d of the main text. 
}
\label{fig4}
\end{figure}

In this section we present additional data supporting that the estimate for the velocity of sound of the $A_{1g}$ mode shows a minimum versus $U/t$ at half filling. Figure~\ref{fig4} shows the velocity of sound of the $A_{1g}$ mode as a function of $U/t$ at half filling ($\delta=0$) and for different temperatures. The curves show a minimum versus $U/t$ (filled symbols). The position of the minima are shown as red squares in Figure 1 of the main text.

\end{document}